\documentclass[runningheads]{llncs}

\usepackage{eccv}

\usepackage{eccvabbrv}

\usepackage{graphicx}
\usepackage{booktabs}

\usepackage[accsupp]{axessibility}  

\usepackage{hyperref}
\usepackage{graphicx,caption}
\usepackage{lipsum}
\usepackage{tikz}
\usepackage{pgfplots}
\pgfplotsset{compat=1.17}
\usepgfplotslibrary{groupplots}
\usepackage{orcidlink}
\usepackage{tabulary}
\usepackage{multirow}
\usepackage{xifthen}
\usepackage{wrapfig}
\newcommand{\N}{AMSNet}
\DeclareMathOperator*{\argmin}{arg\,min}
\newcommand{\tinyFig}[3]{%
\begin{subfigure}[t]{#1\textwidth}
    \centering
    \captionsetup{justification=centering}
    \fontsize{8}{8}
    \tiny
    \begin{tikzpicture}
        \node[anchor=south west, inner sep=0] at (0,0) {\includegraphics[width=\linewidth]{#2}};
    \end{tikzpicture}
    
    \caption{\label{#2}#3}
\end{subfigure}
}
\newcommand{\tinyFigE}[6]{%
    \begin{subfigure}[t]{#1\textwidth}
        \centering
        \tiny
        \captionsetup{justification=centering}
        \fontsize{8}{8}
        \begin{tikzpicture}
            \node[anchor=south west, inner sep=0] at (0,0) {\includegraphics[width=\linewidth]{#2}};
            \ifthenelse{\isempty{#4}}{}{
                \node[anchor=south west, inner sep=0] at (#5) {\textcolor{#6}{\textbf{#4}}}; 
                }
        \end{tikzpicture}
        \caption{\label{#2}#3}
    \end{subfigure}
}
\usepackage{orcidlink}
\begin{document}

\title{Asymmetric Mask Scheme for Self-Supervised Real Image Denoising} 

\titlerunning{AMSNet}

\author{Xiangyu Liao\inst{1} \and
Tianheng Zheng\inst{1} \and
Jiayu Zhong\inst{1} \and
Pingping Zhang\inst{2}\and
Chao Ren\inst{1}\orcidlink{0000-0002-5347-2728}\thanks{Corresponding author.}}

\authorrunning{X.Liao et al.}

\institute{Sichuan University \and
Dalian University of Technology \\
\email{chaoren@scu.edu.cn}}

\maketitle

\begin{abstract}
    In recent years, self-supervised denoising methods have gained significant success and become critically important in the field of image restoration. Among them, the blind spot network based methods are the most typical type and have attracted the attentions of a large number of researchers. Although the introduction of blind spot operations can prevent identity mapping from noise to noise, it imposes stringent requirements on the receptive fields in the network design, thereby limiting overall performance. To address this challenge, we propose a single mask scheme for self-supervised denoising training, which eliminates the need for blind spot operation and thereby removes constraints on the network structure design. Furthermore, to achieve denoising across entire image during inference, we propose a multi-mask scheme. Our method, featuring the asymmetric mask scheme in training and inference, achieves state-of-the-art performance on existing real noisy image datasets. Code will be available at \href{https://github.com/lll143653/amsnet}{https://github.com/lll143653/amsnet}.
\end{abstract}
\section{Introduction}
\label{sec:intro}
Obtaining higher quality images is a key goal within the fields of computer vision \cite{llrr,Zeng_2019_ICCV}. Removing noise significantly enhances image quality, offering improved visual appeal and more efficient processing. Currently, the application of deep learning for image denoising has shown remarkable effectiveness. However, most of these methods rely on paired datasets, which are synthesized from a large number of clean images \cite{anwar2019real,yu2019deep,zamir2022restormer}. These methods have limited applicability to real-world tasks that lack paired data. To overcome this challenge, a series of real paired datasets have been introduced, such as SIDD \cite{abdelhamed2018high} and NIND \cite{brummer2019natural}. Applying supervised denoising methods trained on these real paired datasets yields better performance in handling real-world tasks \cite{zamir2022restormer,ren2021adaptive}.








\definecolor{ht}{RGB}{252, 240, 225}
\begin{figure}[tp]
      \centering
      \tiny
      \tinyFig{0.22}{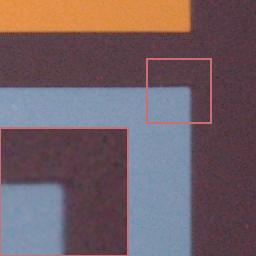}{\textit{GT}}
      \tinyFig{0.22}{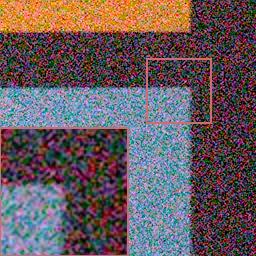}{\textit{Noisy}}
      \tinyFigE{0.22}{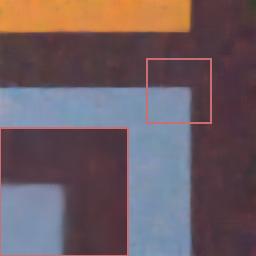}{\textit{AP-BSN}\cite{lee2022apbsn}}{31.24dB}{0.05,2.3}{ht}
      \tinyFigE{0.22}{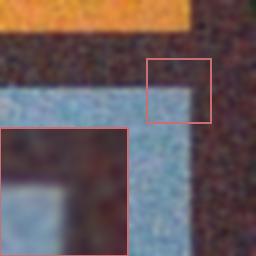}{\textit{CVF-SID}\cite{neshatavar2022cvf}}{26.62dB}{0.05,2.3}{ht}
      \tinyFigE{0.22}{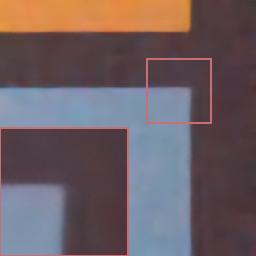}{\textit{LG-BPN}\cite{wang2023lg}}{31.86dB}{0.05,2.3}{ht}
      \tinyFigE{0.22}{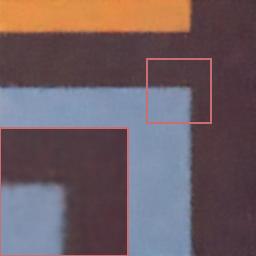}{\textit{BNN-LAN}\cite{li2023spatially}}{31.97dB}{0.05,2.3}{ht}
      \tinyFigE{0.22}{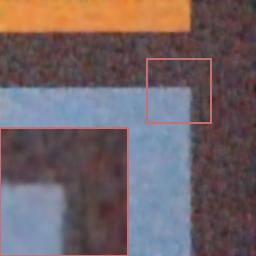}{\textit{SCPGabN}\cite{Lin_2023_ICCV}}{30.35dB}{0.05,2.3}{ht}
      \tinyFigE{0.22}{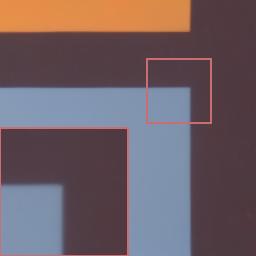}{\textit{\N-P-E}}{33.87dB}{0.05,2.3}{ht}
      \caption{Visual comparison of denoising results on the SIDD validation dataset \cite{abdelhamed2018high} with various methods and our \N\ is able to preserve more details and achieve better visual effects.}
      \label{fig:fig_36}
\end{figure}

However, the complexity of real-world scenarios is highlighted by variations in noise types due to factors such as sensor noise, environmental conditions, and electromagnetic interference. These variations usually result in real datasets struggling to provide complete coverage for real noise types. In response to these challenges, self-supervised denoising methods, which do not require paired samples, have emerged as a promising solution.

Recently, a variety of distinct self-supervised denoising approaches have been proposed. Among them, the Blind Spot Network (BSN), introduced by Noise2Void \cite{krull2019noise2void}, stands out. BSN operates on the assumption that noise is independent and has a zero mean. However, real-world noise usually defies the assumption of independence. To address this discrepancy, AP-BSN \cite{lee2022apbsn} proposes to introduce asymmetric pixel downsampling during training and inference, yielding impressive results in preventing the identity mapping from noisy image to noisy image. Nonetheless, BSN-type strategies that exclude the central pixel to remove noise inevitably lead to a loss of structural information. Moreover, the limitations of BSNs on the networks' receptive field severely restrict its denoiser design.

To address the limitations inherent in denoiser design for BSNs, we draw inspiration from MAE \cite{he2022masked} and introduce a mask-based self-supervised denoising method. This approach overcomes constraints on the denoiser design, and achieves state-of-the-art denoising results via an asymmetric scheme during training and inference. Our contributions can be summarized as follows:
\begin{itemize}
  \item Based on the analysis of typical BSNs, their limitations for network design are provided. To address these limitations, we propose to apply a novel mask strategy to self-supervised denoising tasks. The results validate the versatility of our approach to multiple widely used denoisers, without the network design limitations of BSNs.
  \item Based on the proposed strategy, we design an Asymmetric Mask Scheme based Network (\N) for self-supervised denoising. During the training phase, a single mask scheme is proposed and further optimized by using the proposed mask self-supervised loss. During the inference phase, a multiple masks scheme is applied to complete the denoising of the entire noisy image.
  \item Our method offers an option for the self-supervised denoising methods, allowing for the flexible selection of denoisers. Compared with existing state-of-the-art methods, our method achiecves excellent performance, even on real-world datasets with complex noise patterns.
\end{itemize}
\section{Related Work}
\label{sec:related}
\subsection{Supervised Image Denoising}
The development of convolutional neural networks (CNN) has greatly improved the image denoising tasks \cite{zhang2017beyond,zhang2018ffdnet}. DnCNN \cite{zhang2017beyond} performs favorably against traditional block-based methods \cite{buades2005non,dabov2007image,gu2014weighted} in Gaussian denoising. FFDNet \cite{zhang2018ffdnet} takes a noise level map as input and can handle various noise levels by using a single model. 
However, models trained on additive white Gaussian noise generalize poorly in real scenes due to domain differences between synthetic and real noise. To solve it, CBDNet \cite{guo2019toward} inverts the demosaicing and gamma correction steps in image signal processing (ISP), and then synthesizes signal-dependent Poisson-Gaussian noise in the original space. Zhou et al. \cite{zhou2020awgn} decompose spatially correlated noise into pixel-independent noise through pixel shuffle and then process it using an AWGN-based denoiser. Another approaches are to collect pairs of noisy and clean images to construct real-world datasets. \cite{abdelhamed2018high,plotz2017benchmarking,zamir2022restormer,ren2021adaptive}. Using these real datasets for training, those models are more likely to generalize to the corresponding real noise \cite{anwar2019real,kim2020transfer,liang2021swinir,ren2021adaptive,wang2022uformer,ren2022enhanced}. However, obtaining real datasets is relatively difficult, and the coverage of scenarios is quite limited.
\subsection{Self-Supervised Image Denoising}
Self-supervised techniques reduce the reliance on paired images by training solely on noisy images. Noise2Void \cite{krull2019noise2void} uses blind spot network to remove noise and Noise2Self \cite{batson2019noise2self} creates input-target pairs via pixel erase. Laine19 \cite{laine2019high} and D-BSN \cite{wu2020unpaired} further optimize the BSN and improve its denoising performance. Self2Self \cite{S2S} adopt dropout strategy randomly to denoise on a single noisy image and Noise2Same \cite{xie2020noise2same}. Blind2Unblind \cite{wang2022blind2unblind} propose new denoising losses for self-supervised training. Neighbor2Neighbor \cite{huang2021neighbor2neighbor} samples the noisy image into two similar sub-images to form a noisy-noisy pair for self-supervised training. CVF-SID \cite{neshatavar2022cvf} can disentangle the clean image, signal-dependent and signal-independent noises from the real-world noisy input via various self-supervised training objectives. AP-BSN \cite{lee2022apbsn} employs an asymmetric pixel downsampling strategy, in addition to using BSN to process real-world sRGB noisy images and achiecves improvement. SDAP \cite{Pan_2023_ICCV} uses Random Sub-samples Generation to improve the performance of BSN effectively. The design of BSNs, which predict the central pixel using its neighbors, necessitates a limited receptive field and may limit performance on images with highly detailed textures \cite{laine2019high,wu2020unpaired,wang2023lg,wang2022blind2unblind}.

\section{Proposed Method}
\label{sec:method}
In this section, we perform a detailed analysis of the structural defects of the typical BSNs and propose our self-supervised asymmetric mask scheme. \cref{sec:reap} addresses the structural limitations of BSNs. In \cref{sec:usemask}, we analyze our mask based self-supervised denoising training scheme. In \cref{sec:maksNet},
we provide our multi mask denoising scheme for entire image denoising . In \cref{sec:analy}, we conduct further analysis and propose enhancements to improve the denoising performance.
\subsection{Revisiting BSNs}
\label{sec:reap}
BSN-based methods are typical self-supervised approaches for single image denoising tasks and previous works \cite{krull2019noise2void,batson2019noise2self,wu2020unpaired} have typically assumed that the noise is zero-mean and pixel-wise independent. The optimization of a BSN minimizes the following loss function $\mathcal{L}_{BSN}$ with noisy image $I_N$:
\begin{equation}
  \begin{split}
    \mathcal{L}_{BSN}&=\parallel B(I_N)-I_N\parallel_1
  \end{split}
  \label{math:apbsn}
\end{equation}
where $B$ represents the denoising model within the BSN framework.

\begin{figure}[tp]
  \centering
  \includegraphics[width=0.85\textwidth]{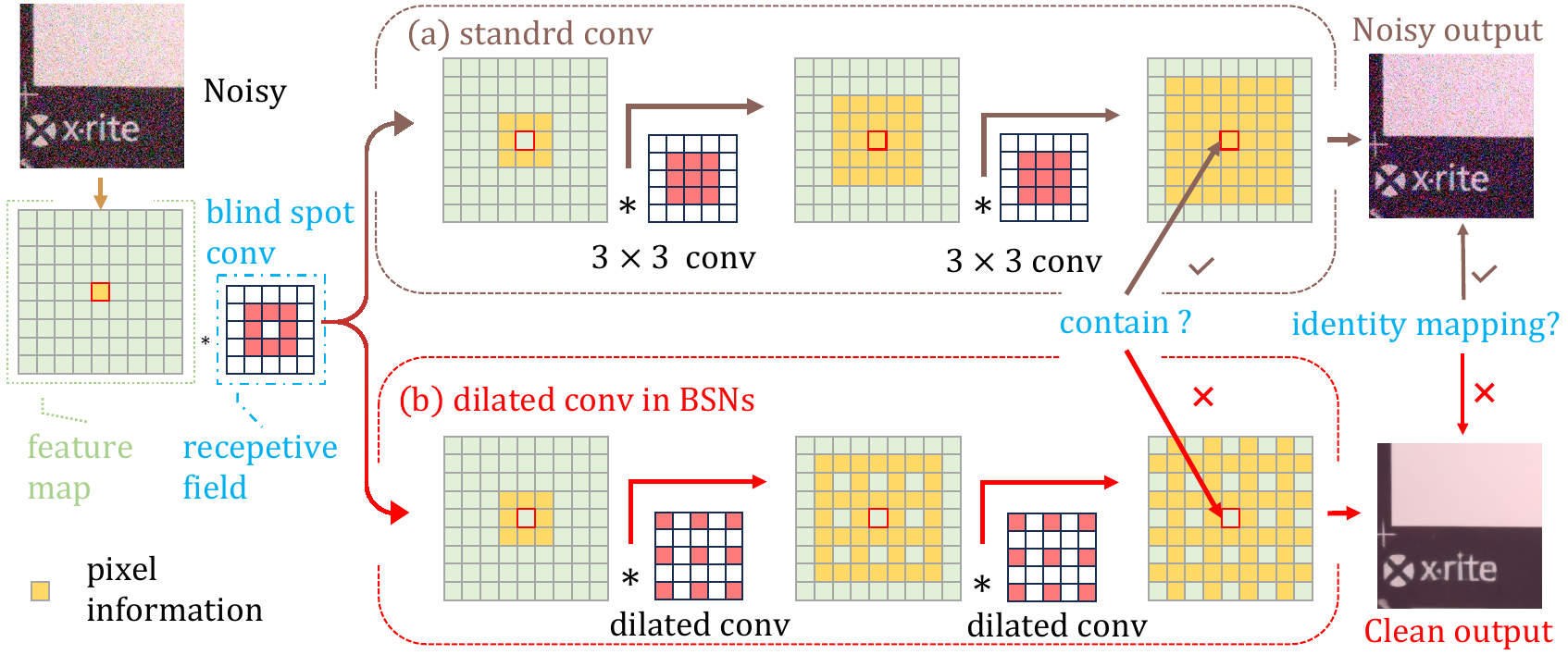}
  \caption{The effect of different receptive fields after blind spot convolution on the final denoising result.}
  \label{fig:bsninfo}
\end{figure}

If $B$ is not constrained in design, minimizing the \cref{math:apbsn} tends to guide the network to produce an output that resembles the original input, essentially resulting in an identity mapping from the noisy image $I_N$ to it self.
\definecolor{seagreen}{RGB}{237, 251, 227}
To circumvent this issue, as illustrated in \cref{fig:bsninfo}, some BSN-based methods \cite{lee2022apbsn,huang2021neighbor2neighbor,Pan_2023_ICCV,wang2023lg,wang2022blind2unblind} utilize blind-spot convolution and introduce restricted operation like dilated convolutions to limit the receptive field and prevent the influence of the input pixel on corresponding output pixel. The yellow patches represent areas containing the original central pixel information, while the green patches denote regions that are not included. The convolution receptive field are highlighted in red. Experiments corresponding to (a) and (b) can be found in \cref{sec:ablation1} that when the convolutional receptive field after blind spot convolution is no longer restricted, the input original pixels are restored.

By employing blind-spot convolution, the central pixel is erased but its information remain contained in its neighborhood pixels. Consequently, the identity mapping occurs if any of the convolutions in subsequent operations have an unconstrained receptive field. To mitigate the tendency toward identity mapping, dilated convolutions or other receptive field-limiting strategies are applied to further isolate the pixel informations out the restore process of corresponding output pixel. With the limited receptive fields, the output pixels is restored by surrounding relevant pixels solely and the noise is eliminated. More details about previous in the \cite{krull2019noise2void,wang2023lg,wu2020unpaired}. Nevertheless, such strategies can compromise the ability of networks to perceive information, potentially diminishing the overall performance \cite{wu2020unpaired}. Moreover, these restrictive techniques can limit network design flexibility, impeding the direct application of advanced denoisers in BSN-based self-supervised frameworks \cite{krull2019noise2void,Pan_2023_ICCV,huang2021neighbor2neighbor}.

Inspired by the work of Masked AutoEncoders (MAE) \cite{he2022masked}, we observe that complete image reconstruction is still achievable from the remaining areas, even when parts of the image are masked. Based on this contemplation, we ask whether using masking can address the design limitations of BSNs. Upon thorough analysis and careful design, we propose an mask-based self-supervised denoising method that operates effectively without the limitations typically associated with BSN-based architectures.

\begin{figure}[t]
  \centering
  \includegraphics[width=0.85\textwidth]{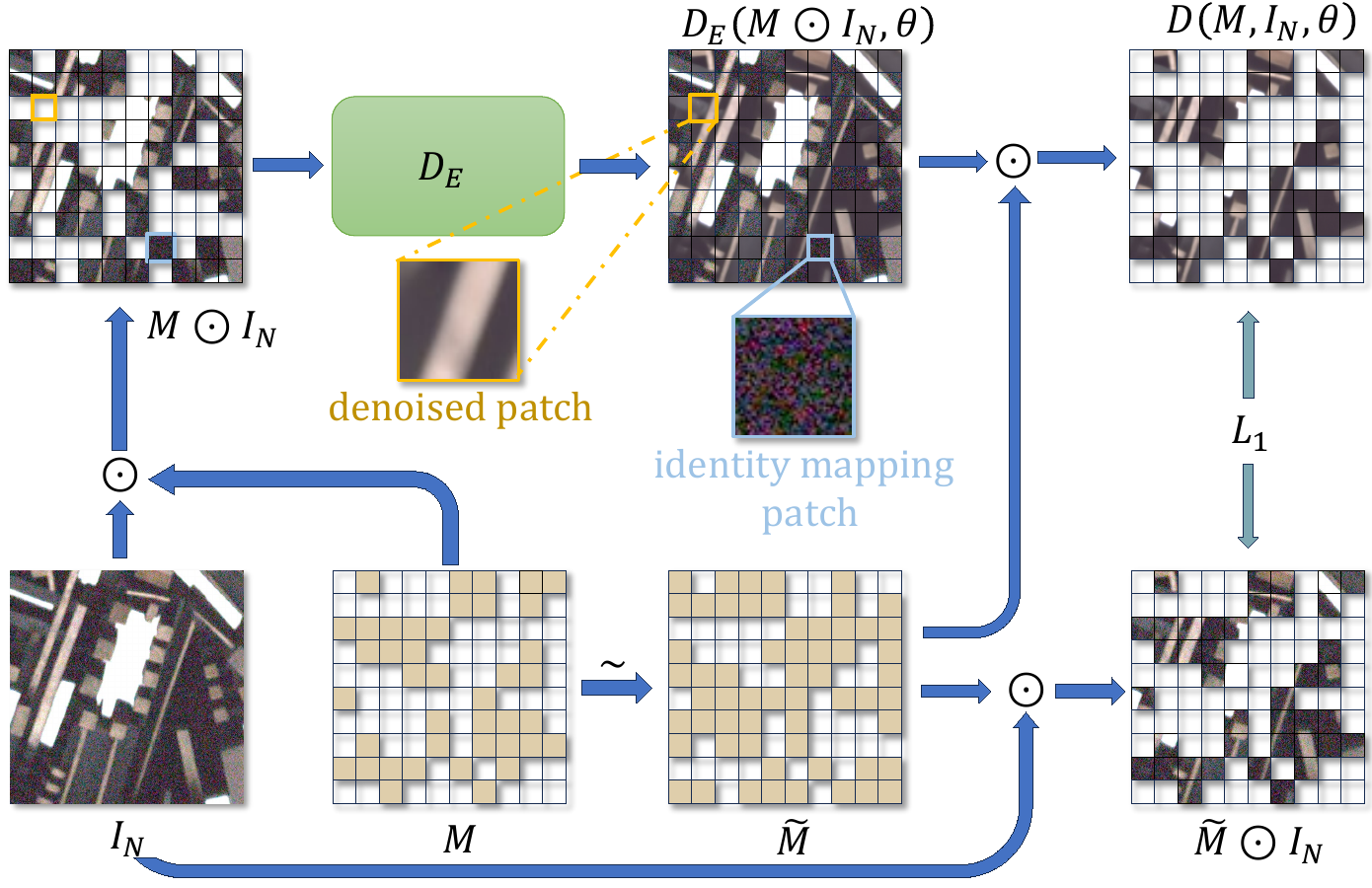}
  \caption{Denoising via mask matrix $M$ and the noisy image $I_N$.}
  \label{fig:mask_denoise}
\end{figure}

\subsection{Training via Single Mask Scheme}

\label{sec:usemask}

In \cref{sec:reap}, our analysis necessitates the isolation of original pixel information out the restoration to preclude identity mapping during the self-supervised training phase. Contrary to BSN-based approaches, which isolate pixels information using blind-spot and dilated convolutions, our approach masking original pixels at the input phase then barring it from the restoration workflow. Thus there is no need to restrict the design of the receptive field, and it also avoids the incorporation of original pixel information into the restoration process.

As illustrated in \cref{fig:mask_denoise}, the masked pixels within the masked areas are reconstructed exclusively from the surrounding unmasked pixels. This approach effectively eliminates noise in these areas and prevents identity mapping. In contrast, the unmasked areas undergo identity mapping, as their pixel information remains unchanged during the restoration process.

A comprehensive depiction of our masking denoising process is illustrated in \cref{fig:mask_denoise}. Let $I_N\in R^{c\times h\times w}$ be a single noisy image, where $c$ represents the number of channels, $h$ and $w$ denote the height and width of image. We randomly mask some pixels via a $0,1$ mask matrix $M$, where $M\in R^{c\times h\times w}$. That means these pixels are replaced with zero tokens. By experience, we mask about 50\% noisy pixels. Consequently, the masked image is given by $M\odot I_N$, where the $\odot$ indicates element-wise multiplication. Then, the masked image $M\odot I_N$ is fed into the denoiser $D_E$, resulting in a restored image $D_E(M\odot I_N,\theta)$, where $\theta$ represents parameters of the denoiser $D_E$.

Based on previous analysis, we can extract the denoised pixels at corresponding positions from the output of the denoiser $D_E$ using the following equation:
\begin{equation}
  \begin{split}
    D(M,I_N,\theta)=\tilde{M}\odot D_E(M\odot I_N,\theta)
  \end{split}
  \label{equa:D}
\end{equation}
where $D$ represents the entire denoising process, which includes masking with $M$ and extracts the final non-identity mapping restoration result with the inverse matrix $\tilde{M}$. The $\tilde{M}$ is the complement of $M$ and indicates to the restoration pixels. Building upon BSN, our optimization process can be represented as follows:

\begin{equation}
  \begin{split}
    \argmin_{\theta}\parallel \tilde{M}\odot D_E(M\odot I_N,\theta)-\tilde{M}\odot I_N \parallel_1
  \end{split}
\end{equation}
where we use the $L_1$ norm for denoiser optimization. Our previous equation also adhere the assumption that the noise is zero-mean and pixel-wise independent. More details about optimization are available in the Supplementary Materials.

\begin{figure}[t]
  \centering
  \includegraphics[width=0.85\textwidth]{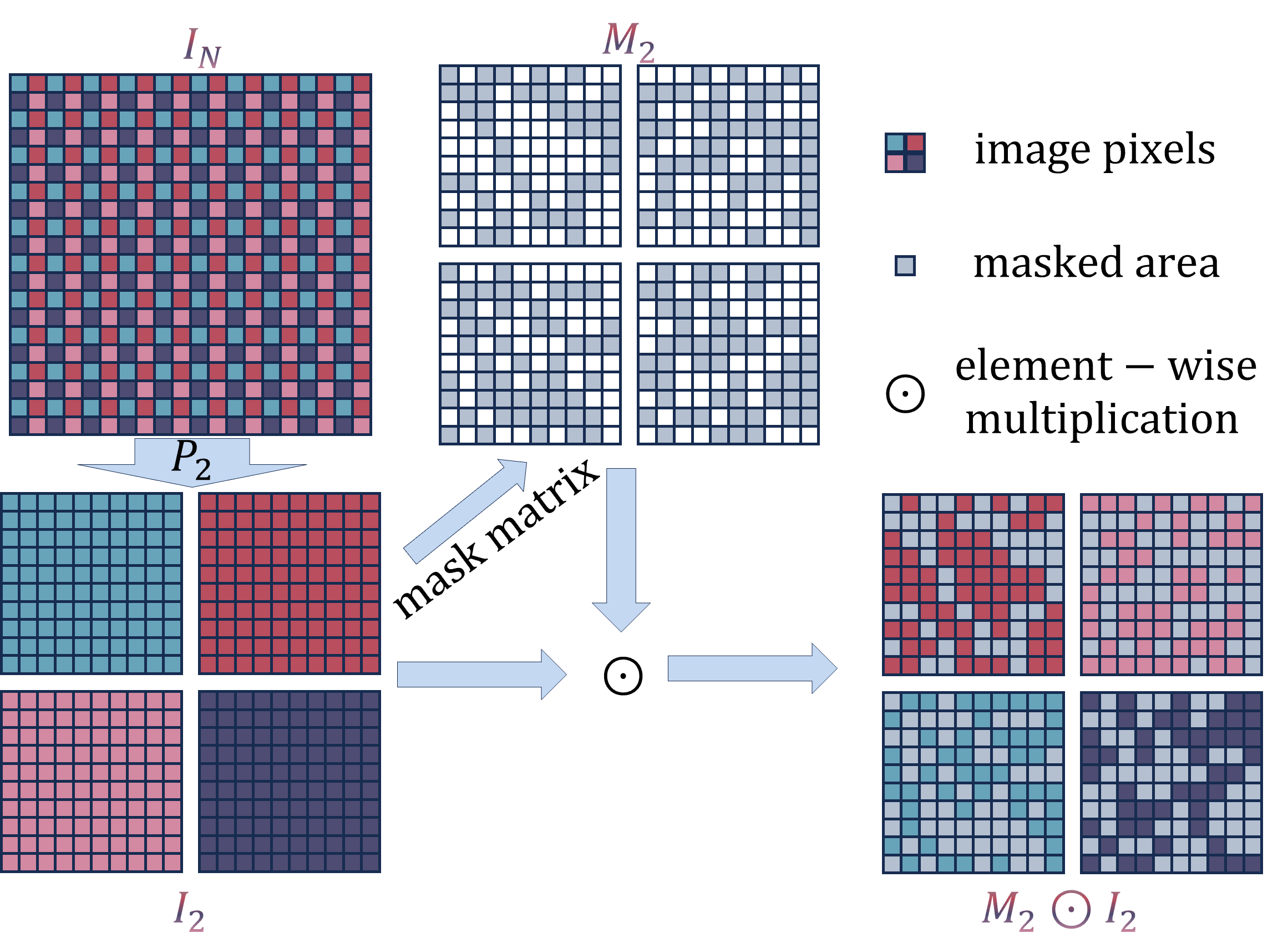}
  \caption{Pixel downsamplinga and mask. Here we take the $P_2$ as an example.}
  \label{fig:subMask}
\end{figure}

\begin{figure*}[t]
  \centering
  \includegraphics[width=0.99\textwidth]{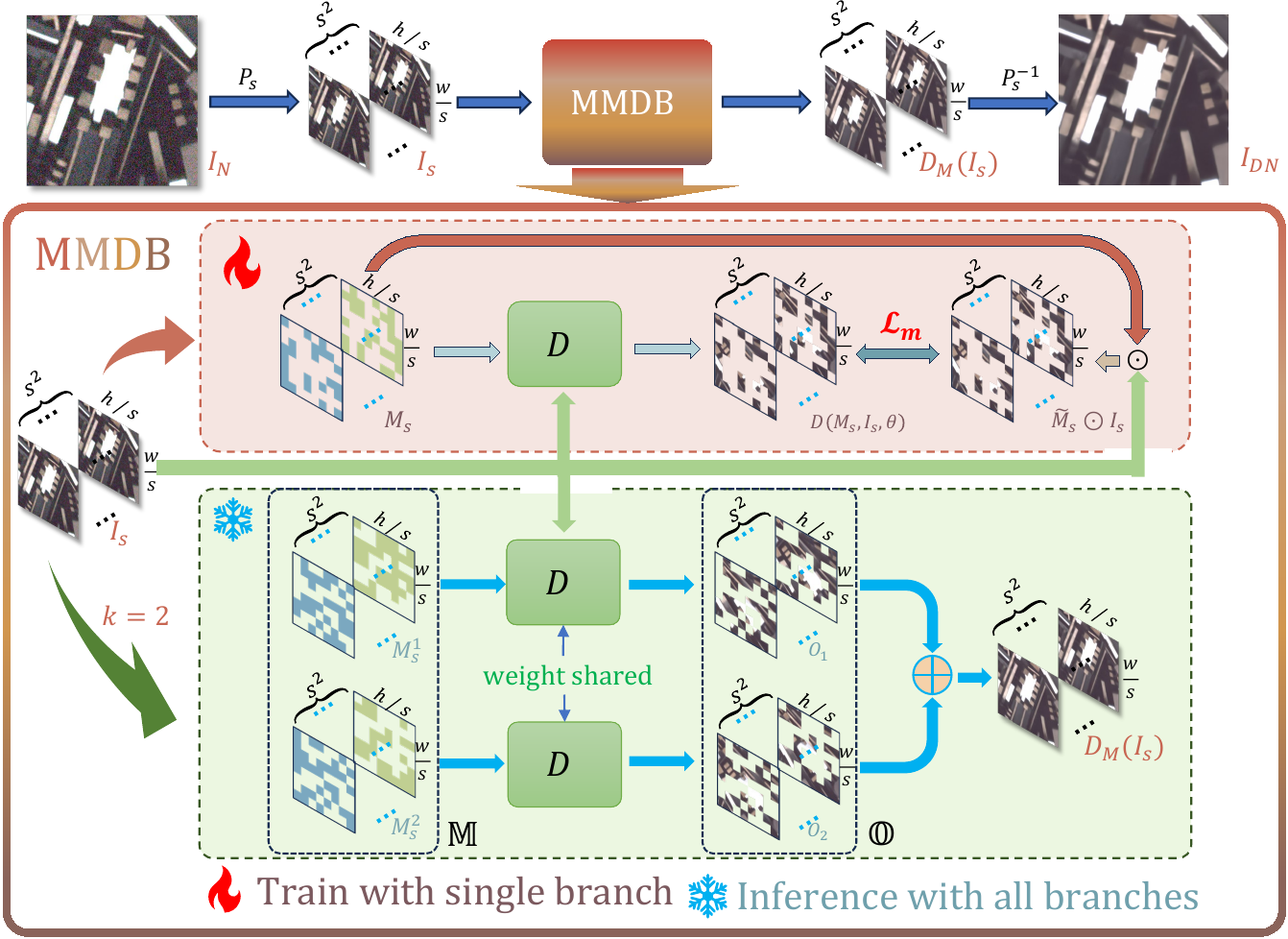}
  \caption{\textbf{Overview of the proposed asymmetric mask scheme}. The $D$ is depicted in \cref{fig:mask_denoise}, which takes sub-samples set $I_s$ of the noisy image and the corresponding binary mask matrices set $M_{s}$ to denoise the specified regions. Here, \textbf{we present a configuration with 2 denoising branches as an example}. During the training phase, a single branch is employed for optimization of denoiser. During the inference phase, we utilize all branches to derive restoration results for entire noisy image.}
  \label{fig:networkStructure}
\end{figure*}

However, in real-world noisy scenarios, noise usually deviates from the assumption of spatial independence. Taking inspiration from AP-BSN \cite{lee2022apbsn}, we introduce a pixel downsampling (PD) strategy, denoted as $P_s$ with a stride factor of $s$. The $P_s$ operation disrupts the spatial correlation among the noise, ensuring that the resulting sub-samples align with our assumptions, as depicted in \cref{fig:subMask}.
The training scheme of real-world noisy scenarios is illustrated as \cref{fig:networkStructure}. Upon applying $P_s$ operation on real world noisy image $I_N$, we obtain a set of sub-samples, which represent as $I_s=\{I_{sub,1},\dots,I_{sub,s^2}\}$, where $I_{sub,k}\in R^{c\times \frac{h}{s}\times \frac{w}{s}}$ and $I_s=P_s(I_N)$. Independent masking is then performed on each sub-sample  using a consistent probability distribution, represented by $M_s\odot I_s$, where $M_s$ is the aggregate set of binary mask matrices for all sub-samples, i.e., $M_s=\{M_{sub,1},\dots,M_{sub,s^2}\}$, and each $M_{sub,k}\in R^{c\times \frac{h}{s}\times \frac{w}{s}}$. These masked sub-samples set can then be processed by the denoiser $D_E$.

The overall optimization process can be expressed as the minimization of the following mask self-supervised loss:
\begin{equation}
  \begin{split}
    \mathcal{L}_{m}(M_s,I_s)=\parallel \tilde{M_s}\odot (D_E(M_s\odot I_s,\theta)- I_s)\parallel_1
  \end{split}
  \label{math:loss}
\end{equation}

With $\mathcal{L}_{m}(M_s, I_s)$, we successfully remove real world noise from the masked pixels that $M_s$ indicated and free the denoiser limitations inherent in BSNs.


\subsection{Inference via Multi Mask Scheme}

\label{sec:maksNet}
To enable comprehensive denoising across the entire image, we devise an Asymmetric Mask Scheme based Network (\N), which implemente a Multi-branch Mask complementary Denoising Block (MMDB).

The overview of our scheme is depicted in \cref{fig:networkStructure}. Our approach, MMDB, utilizes $k$ denoising branches, where $k \geq 2$, to restore the entire image. The denoising branches are equivalent to the entire denoising process in \cref{fig:mask_denoise}. First, we create a sub-samples set $I_s$ from the original noisy image $I_N$ with $P_s$, which helps distribute the spatial correlation of noise. For each denoisng branch, MMDB generates a series of mask matrices set $M_s$, denoted as $\mathbb{M}={M_s^1,\dots,M_s^k}$, ensuring that the coverage areas across branches are unique and non-overlapping while maintaining a roughly equivalent coverage proportion. The total coverage of these masks spans all pixels, as indicated by the equations 
\begin{equation}
  \sum_{i=1}^kM_s^i=(k-1)\mathbb{I} ,\sum_{i=1}^k\tilde{M}_s^i=\mathbb{I}
\end{equation}
where $\mathbb{I}$ refers to a matrix has same shape to $M_s^i$ and all elements are one. More details in Supplementary Materials.

Each branch utilizes the same denoiser $D_E$ and the output of all branches is denoted as $\mathbb{O}=O_1,\dots,O_k$. For a given branch $i$, where $1 \leq i \leq k$, the $I_s$ and corresponding $M^{i}_{s}$ are combined to produce the output $O_i$ independently:
\begin{equation}
  O_i= D_i(M^{i}_{s}, I_s,\theta)
\end{equation}
where $D_i$ is the $i$-th denoising branch and $\theta$ is the parameter of denoiser $D_E$. The output of MMDB can be represented as the output sum of all branches:
\begin{equation}
  \begin{split}
    D_M(I_s)&=\sum_{i=1}^{k}O_i=\sum_{i=1}^{k}D_i(M^{i}_{s}, I_s,\theta)=\sum_{i=1}^{k}\tilde{M}^{i}_{s}\odot D_E(M^{i}_{s}\odot I_s,\theta)
  \end{split}
\end{equation}
where, $D_M(I_s)$ refer to the output of MMDB. In the end, we can fully describe our framework as follows:
\begin{equation}
  \begin{split}
    I_{DN}&=P^{-1}_s(D_M(P_s(I_N))) \\
  \end{split}
  \label{math:output}
\end{equation}
where $P^{-1}_s$ represents the inverse operation of pixel downsampling $P_s$ with stride $s$. More details about inference are in Supplementary Materials.

\subsection{Analysis and Enhancement}

\label{sec:analy}
Based on our insights, our approaches achieves self-supervised noise removal. We introduce an asymmetric training-inference scheme that not only minimizes optimization costs but also ensures comprehensive denoising during inference. During the training phase, we use a single branch to restore a portion of the pixels with the loss function $\mathcal{L}_m$ for expedited optimization of $D_E$. During the inference phase, all denoising branches collaborate to restore all noisy pixels, thus achieving denoising for the entire image. The innovative use of mask releases us from the restrictive structural demands of blind spot networks, thereby expanding our choose of advanced denoisers and provides new possibilities for the design of the loss function.

Despite these achievements, we still encounter certain challenges. To meet the assumption of noise independence, we have integrated a pixel downsampling (PD) strategy that disrupts the correlation between noise. However, the PD strategy also destroys the structural integrity of the image and causing irreversible information loss. Meanwhile, since the restored pixels are reconstructed from surrounding pixels, there may be minor color shifting. Those approaches inadvertently results in pixel discontinuity within the denoised image, manifesting as checkerboard effect, as exemplified in Supplementary, where the pixels arrangement lacks the smooth continuity of the ground-truth. This greatly affects the final denoising quality and visual performance.
\subsection{Analyzing of Checkerboard and Solutions}

\label{sec:checkerboard}
Compared with ground-truth, the interlaced pixel arrangement in denoised images noticeably reduces smoothness. To address this and promote the generation of higher quality images, we utilize the model \N-B trained via \cref{math:loss} for full image denoising. Subsequently, we introduce a priori smoothness loss $\mathcal{L}_p$ for fine-tuning:
\begin{equation}
  \begin{split}
    \mathcal{L}_p(I)&=\sum_{i,j}\sqrt{(I_{i+1,j}-I_{i,j})^2+(I_{i,j+1}-I_{i,j})^2}
  \end{split}
\end{equation}
where $I$ denotes the image and $i,j$ represent the coordinates of pixel. During fine-tuning, the total loss $\mathcal{L}_t$ which incorporates both the original mask loss $\mathcal{L}_m$ and priori smoothness loss $\mathcal{L}_p$ is used :
\begin{equation}
  \begin{split}
    \mathcal{L}_t&=\lambda\mathcal{L}_p(I_{DN})+\sum_{i=1}^{k}\mathcal{L}_m(M_s^i,I_s) =\lambda\mathcal{L}_p(I_{DN})+\parallel I_{DN}-I_N\parallel_1
  \end{split}
\end{equation}
where we restore all noisy pixels and $\lambda$ is the weight coefficient for the priori smoothness loss and set to 0.01 by experience, $k$ represents the number of denoise branches. The results obtained through using $\mathcal{L}_t$ are labeled as \textbf{P}. Additionally, to further eliminate the checkerboard effect during inference, we introduce the random replacement refinement strategy \cite{lee2022apbsn}. The results obtained through the use of the random replacement refinement during inference are labeled as \textbf{E}. Consequently, the basic model trained by only $\mathcal{L}_m$ is denoted as \textbf{\N-B}, the fine-tuning version with $\mathcal{L}_t$  is denoted as \textbf{\N-P}, and the version further with random replacement refinement for checkerboard suppression during the inference is denoted as \textbf{\N-P-E} and \textbf{\N-B-E}. For more details refer to the Supplementary Materials.

\section{Experiments}

\label{sec:experiments}

\subsection{Experiment Configurations}

\textbf{Dataset.} We conduct experiments on the widely-used SIDD \cite{abdelhamed2018high}, DND \cite{plotz2017benchmarking}, and PolyU \cite{xu2018real} datasets. During the training phase, we utilize the SIDD-MEDIUM dataset, which comprises 320 noisy-clean image pairs, and only the noisy images are used for training. The SIDD validation and benchmark each contain 1280 color images (each with a resolution of $256\times 256$). The DND dataset includes 50 high-resolution noisy images and 1000 sub-images (each with a resolution of $512\times 512$) for benchmark. The PolyU dataset contains 100 real-world noisy-clean image pairs (each with a resolution of $512\times 512$) for validation.

\noindent\textbf{Metric.} To evaluate \N\ and compare with other denoising methods, we introduce peak signal-to-noise ratio (PSNR) and structural similarity (SSIM) metrics.

\noindent\textbf{Denoiser.} We select several classic denoising models as our denoisers, including Restormer \cite{zamir2022restormer}, DeamNet \cite{ren2021adaptive}, DnCNN \cite{zhang2017beyond}, NAFNet \cite{chen2022simple} and UNet \cite{ronneberger2015u}. Throughout this paper, we choose Restormer as our default denoiser.

\noindent\textbf{Implementation.} During the training process, we implemente our model using PyTorch 2.1.0 \cite{paszke2017automatic} and train on an NVIDIA RTX3090 GPU. For optimization, we employ AdamW \cite{paszke2017automatic} with default settings and set the initial learning rate set at 0.0001. More details in Supplementary Materials.
\begin{table*}[htp]
    \centering
    \setlength\tabcolsep{3pt}
    \fontsize{8}{8}\selectfont
    \renewcommand{\arraystretch}{1.4}
    \caption{Quantitative PSNR (dB) / SSIM results on SIDD and DND. Here we exploit Restormer \cite{zamir2022restormer} as our denoiser and set $k=2$. $\mathcal{R}^3$ denotes the random replacement refinement \cite{lee2022apbsn}. By default, we use $P_5$ for training and $P_2$ for inference.}
    \begin{tabular}{>{\centering\arraybackslash}m{1.5cm}>{\centering\arraybackslash}m{1.5cm}c>{\centering\arraybackslash}m{1.8cm}>{\centering\arraybackslash}m{1.9cm}>{\centering\arraybackslash}m{1.8cm}}
        \toprule
        Type of supervision         & Training data                      & Method                                     & SIDD Validation                & SIDD Benchmark       & DND benchmark         \\ \midrule
        \multirow{2}{*}{\shortstack{Non-learning\\  based}} & \multirow{2}{*}{-}                    & BM3D \cite{dabov2007image}                          & 31.75/0.7061                   & 25.65/0.685          & 34.51/0.8507          \\
                                    &                                    & WNNM \cite{gu2014weighted}                 & 26.31/0.5240                   & 25.78/0.809          & 34.67/0.8646          \\ \midrule
        \multirow{6}{*}{Supervised} & \multirow{6}{*}{\shortstack{Paired\\noisy-clean}}   & DnCNN \cite{zhang2017beyond}                        & 26.20/0.4414                   & 23.66/0.583          & 32.43/0.7900          \\
                                    &                                    & CBDNet \cite{guo2019toward}                & 30.83/0.7541                   & 33.28/0.868          & 38.06/0.9421          \\
                                    &                                    & RIDNet \cite{anwar2019real}                & 38.76/0.9132                   & 37.87/0.943          & 39.25/0.9528          \\
                                    &                                    & VDN \cite{yue2019variational}              & 39.29/0.9109                   & 39.26/0.955          & 39.38/0.9518          \\
                                    &                                    & DeamNet \cite{ren2021adaptive}             & 39.40/0.9169                   & 39.35/0.955          & 39.63/0.9531          \\
                                    &                                    & Restormer \cite{zamir2022restormer}        & -                              & 40.02/0.960          & 40.03/0.9560          \\ \midrule
        \multirow{3}{*}{\shortstack{Pseudo\\supervised}}  & \multirow{2}{*}{\shortstack{Unpaired\\noisy-clean}} & GCBD \cite{chen2018image}                           & -                              & -                    & 35.58/0.9217          \\
                                    &                                    & C2N \cite{jang2021c2n}                     & 35.36/0.8901                   & 35.35/0.937          & 37.28/0.9237          \\ \cline{2-6}
                                    & Paired noisy-noisy                 & R2R \cite{pang2021recorrupted}             & 35.04/0.8440                   & 34.78/0.898          & 37.61/0.9368          \\ \midrule
        \multirow{9}{*}{\shortstack{Self\\supervised}}   & \multirow{9}{*}{\shortstack{Single\\noisy}}      & AP-BSN+$\mathcal{R}^3$ \cite{lee2022apbsn} & 36.74/0.8501                   & 36.91/0.931          & 38.09/0.9371          \\
                                   &                                      &CVF-SID \cite{neshatavar2022cvf}           & 34.17/0.8719                   & 34.71/0.917          & 36.50/0.9233          \\
                                    &                                    & LG-BPN+$\mathcal{R}^3$ \cite{wang2023lg}   & 37.31/0.8860                   & 37.28/0.936          & 38.43/0.9423          \\
                                    &                                    & SCPGabN \cite{Lin_2023_ICCV}               & 36.53/0.8860                   & 36.53/0.925          & 38.11/0.9393          \\
                                    &                                    & BNN-LAN \cite{li2023spatially}             & 37.39/0.8830                   & 37.41/0.934          & 38.18/0.9386          \\
                                    &                                    & \textbf{\N-B}                              & \textbf{36.60/0.8551}          & \textbf{36.60/0.917} & \textbf{38.12/0.9377} \\
                                    &                                    & \textbf{\N-P}                              & \textbf{36.71/0.8519}          & \textbf{36.66/0.918} & \textbf{38.24/0.9379} \\
                                    &                                    & \textbf{\N-B-E}                            & \textbf{36.73/0.8932}                              & \textbf{37.65/0.939}                 &            \textbf{38.56/0.9402}           \\
                                    &                                    & \textbf{\N-P-E}                            & \textbf{37.93}/\textbf{0.8948} & \textbf{37.87/0.941} & \textbf{38.70/0.9471} \\ \bottomrule
    \end{tabular}
    \label{table:Quantitative}
\end{table*}









\definecolor{ht}{RGB}{252, 240, 225}
\begin{figure*}[tp]
      \centering
      \tiny
      \tinyFig{0.11}{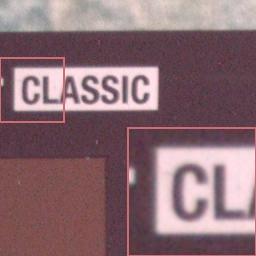}{}
      \tinyFig{0.11}{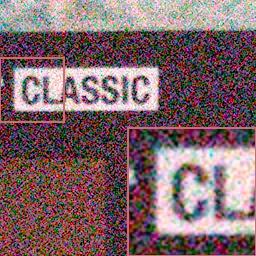}{}
      \tinyFigE{0.11}{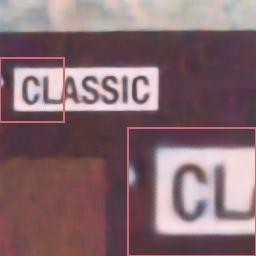}{}{29.47dB}{0.13,1.1}{ht}
      \tinyFigE{0.11}{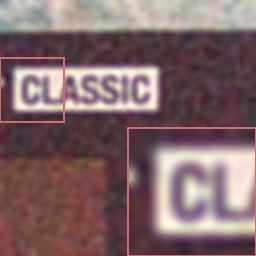}{}{24.98dB}{0.13,1.1}{ht}
      \tinyFigE{0.11}{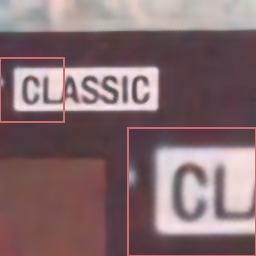}{}{30.10dB}{0.13,1.1}{ht}
      \tinyFigE{0.11}{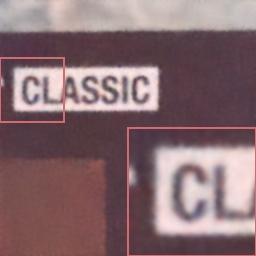}{}{30.34dB}{0.13,1.1}{ht}
      \tinyFigE{0.11}{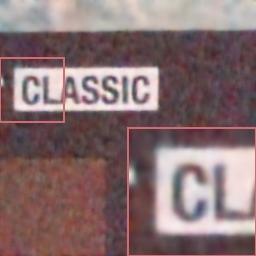}{}{29.06dB}{0.13,1.1}{ht}
      \tinyFigE{0.11}{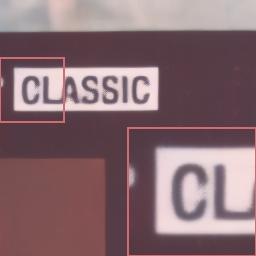}{}{30.96dB}{0.13,1.1}{ht}
      \caption{Visual comparison of our method against other denoising methods on the SIDD validation dataset. The image is (a) \textit{GT} (b) \textit{Noisy} (c) \textit{AP-BSN}\cite{lee2022apbsn} (d) \textit{CVF-SID}\cite{neshatavar2022cvf} (e) \textit{LG-BPN}\cite{wang2023lg} (f) \textit{BNN-LAN}\cite{li2023spatially} (g) \textit{SCPGabN}\cite{Lin_2023_ICCV} (h) \textit{\N-P-E}.}
      \label{fig:fig_47}
\end{figure*}









\begin{figure}[tp]
      \centering
      \tiny
      \tinyFig{0.12}{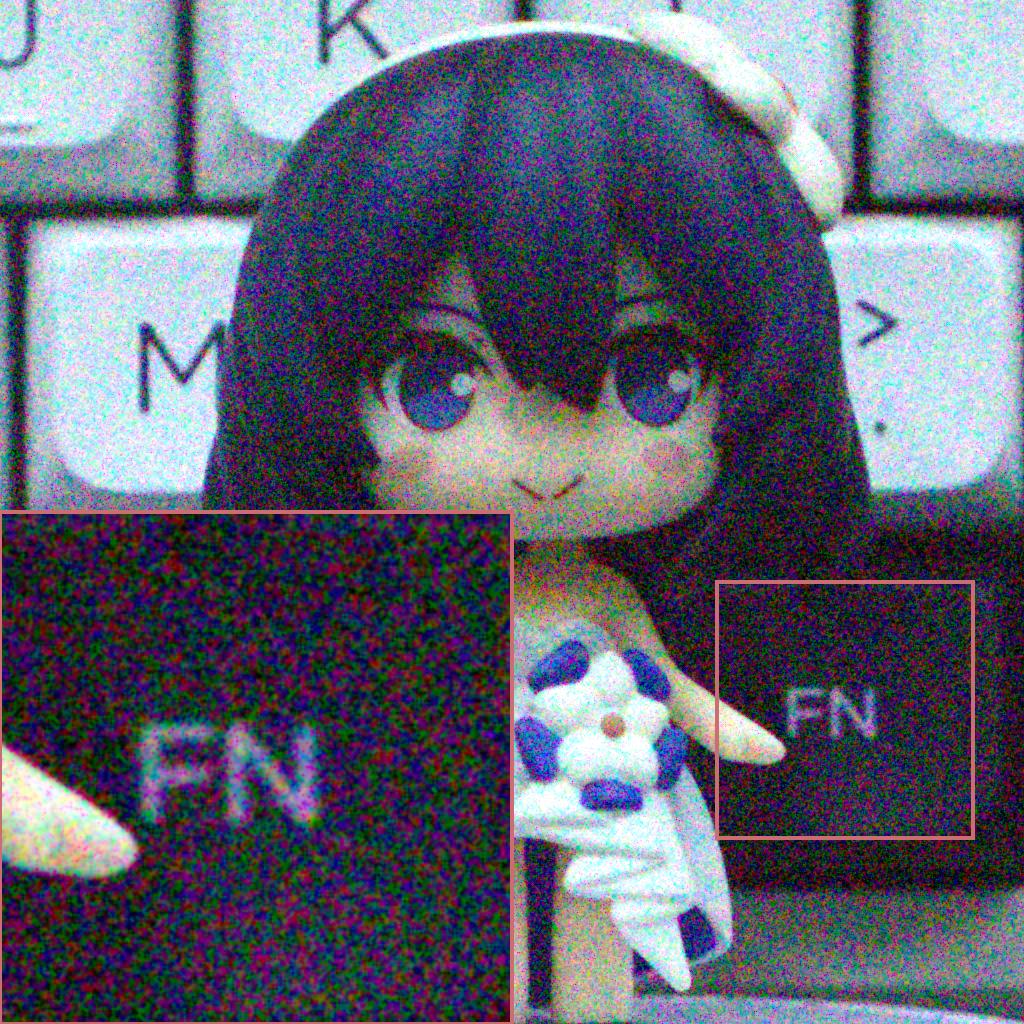}{}
      \tinyFig{0.12}{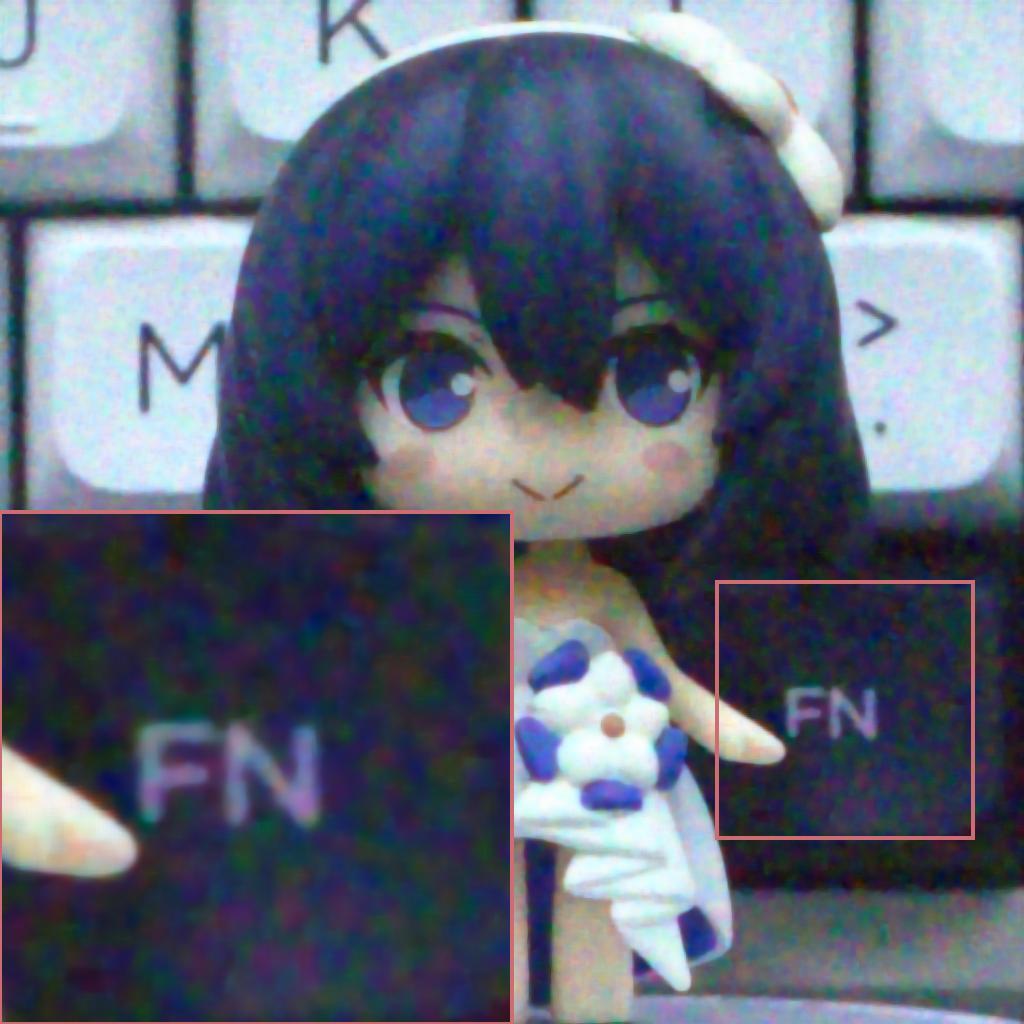}{}
      \tinyFig{0.12}{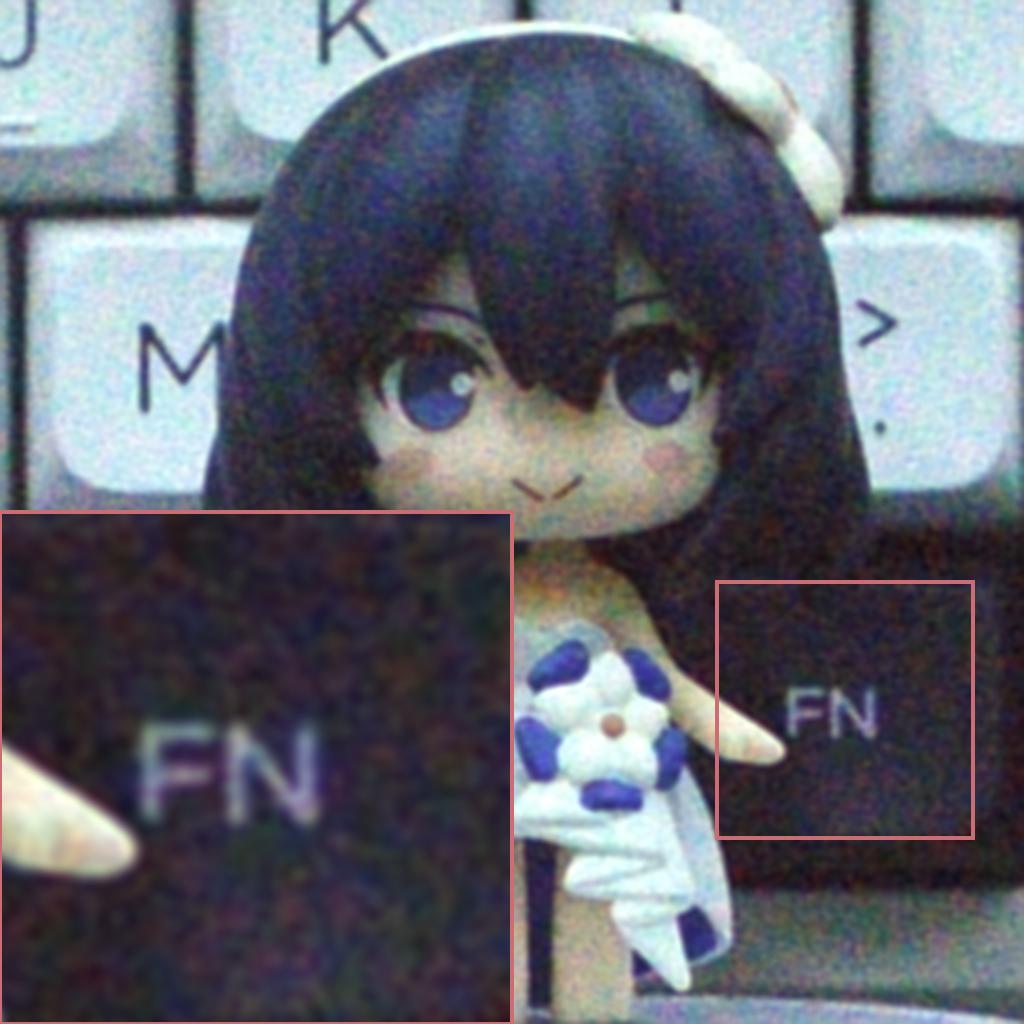}{}
      \tinyFig{0.12}{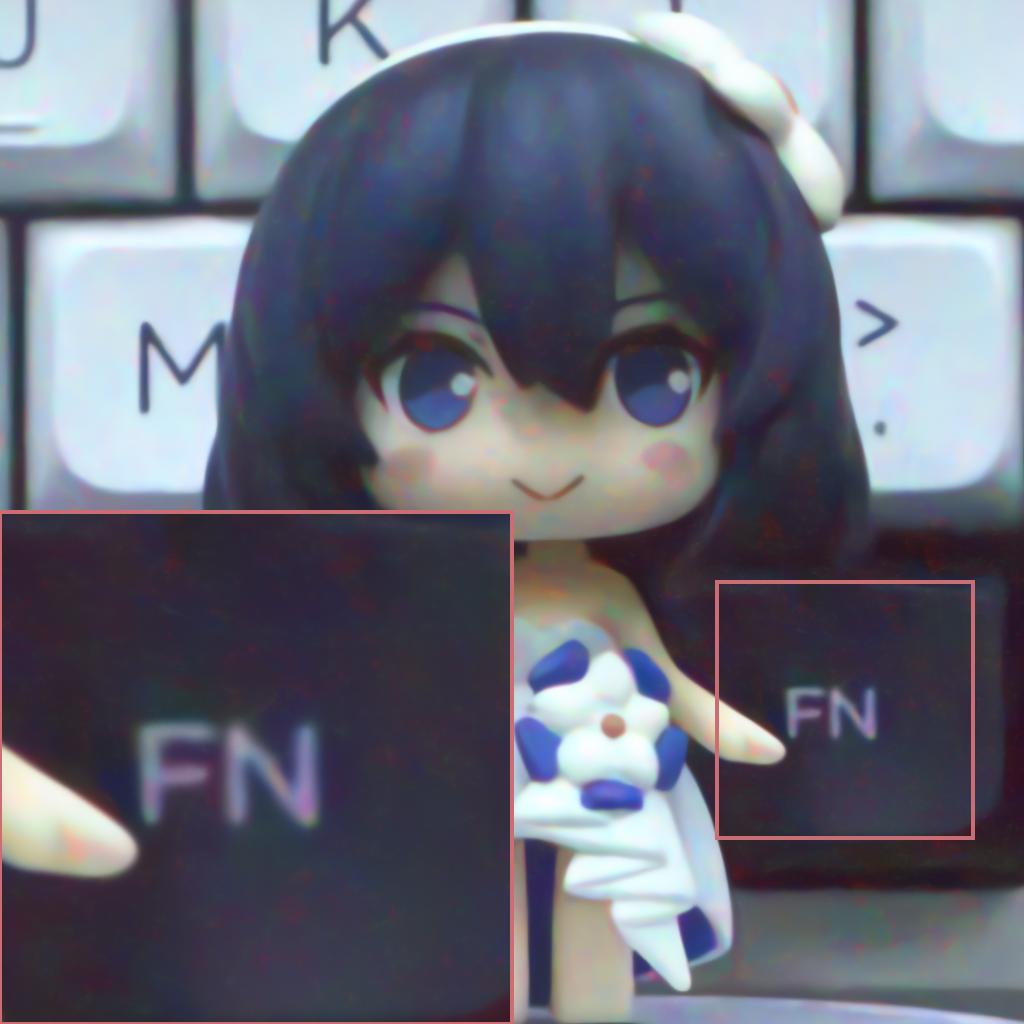}{}
      \tinyFig{0.12}{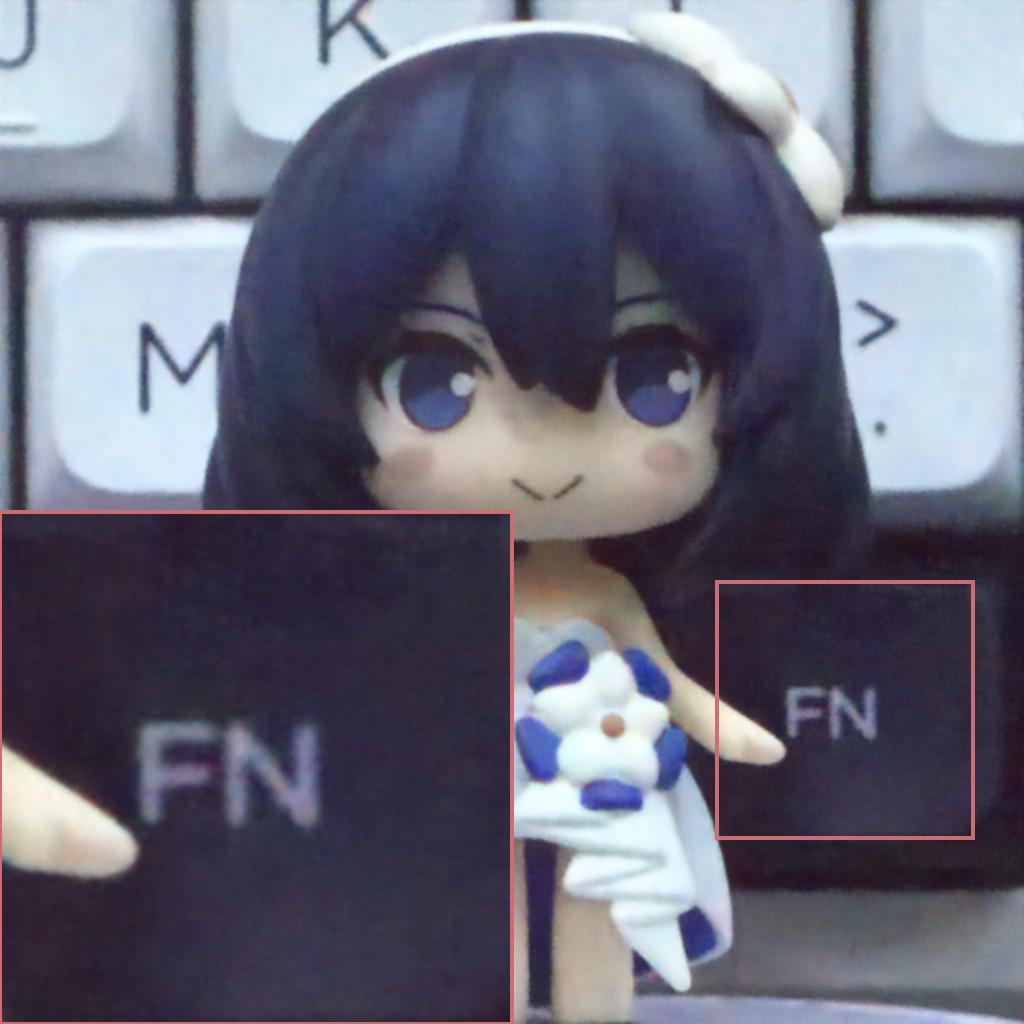}{}
      \tinyFig{0.12}{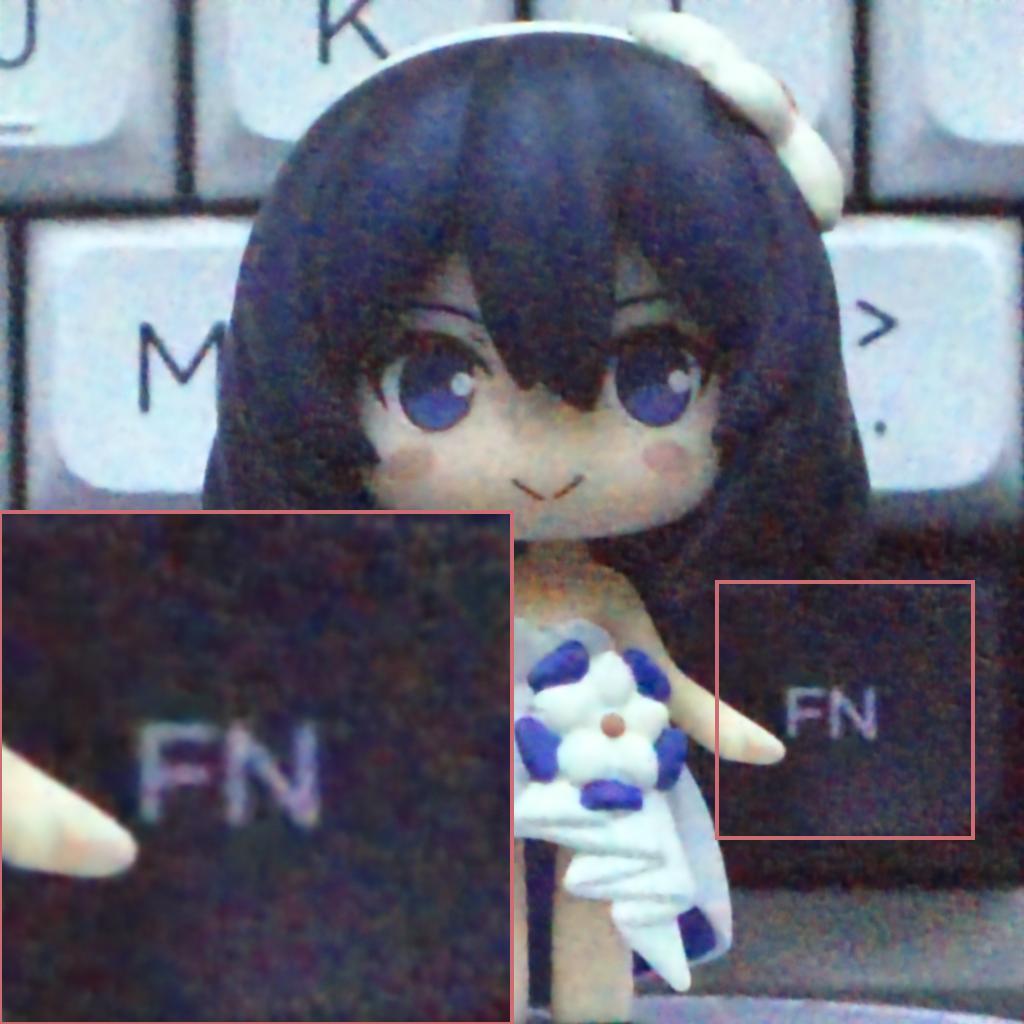}{}
      \tinyFig{0.12}{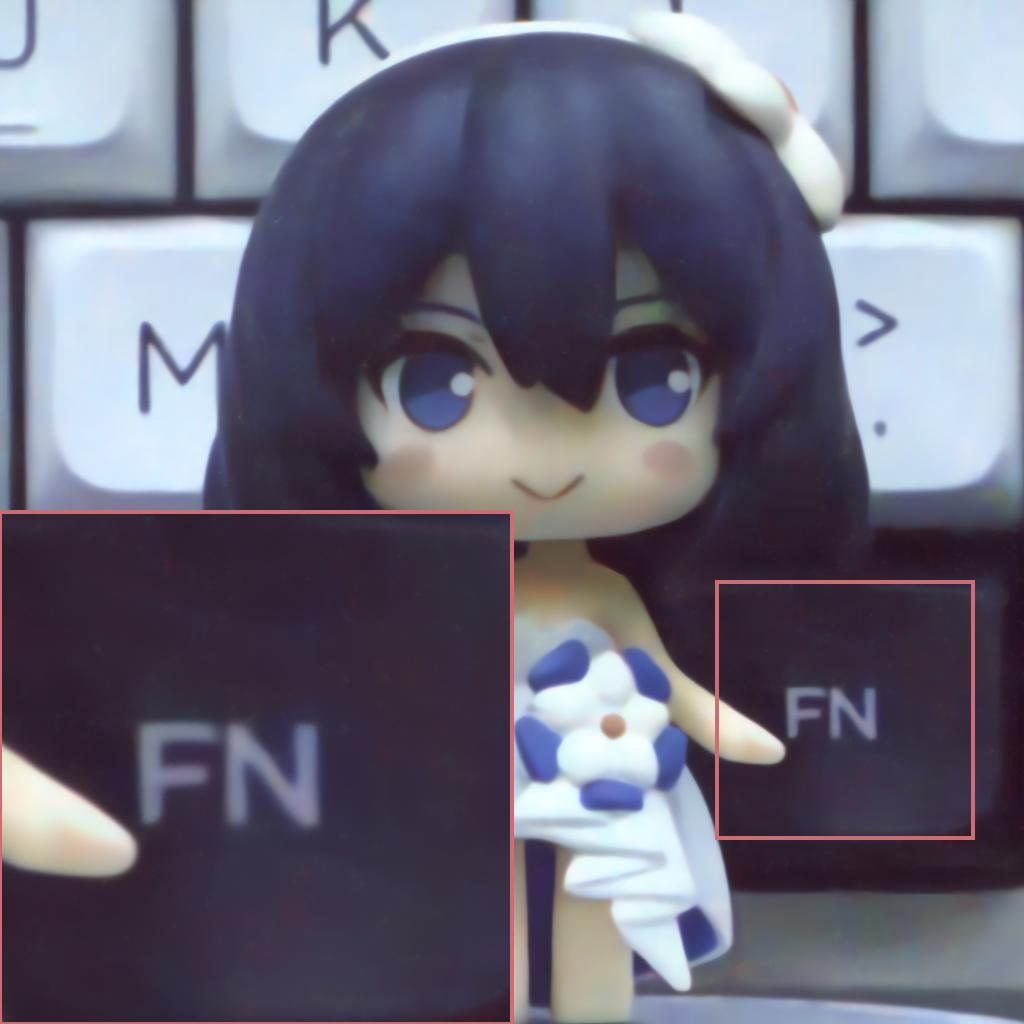}{}
      \caption{The result of denoising noise images taken with Canon EOS M5 camera. The image is (a) \textit{Noisy} (b) \textit{AP-BSN}\cite{lee2022apbsn} (c) \textit{CVF-SID}\cite{neshatavar2022cvf} (d) \textit{LG-BPN}\cite{wang2023lg} (e) \textit{BNN-LAN}\cite{li2023spatially} (f) \textit{SCPGabN}\cite{Lin_2023_ICCV} (g) \textit{\N-P-E}.}
      \label{fig:canon_fig_0}
\end{figure}








\definecolor{ht}{RGB}{252, 240, 225}
\begin{figure}[h]
      \centering
      \tiny
      \tinyFig{0.12}{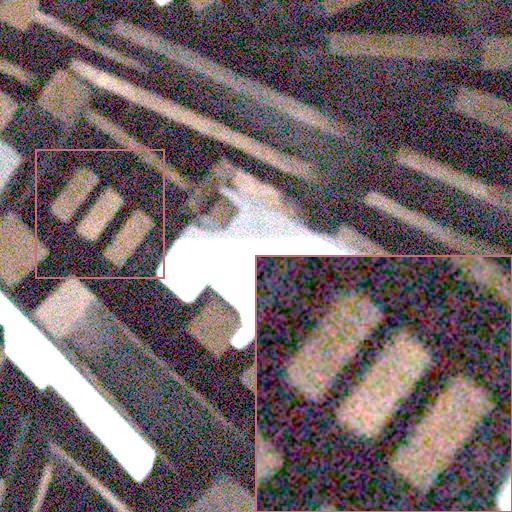}{}
      \tinyFigE{0.12}{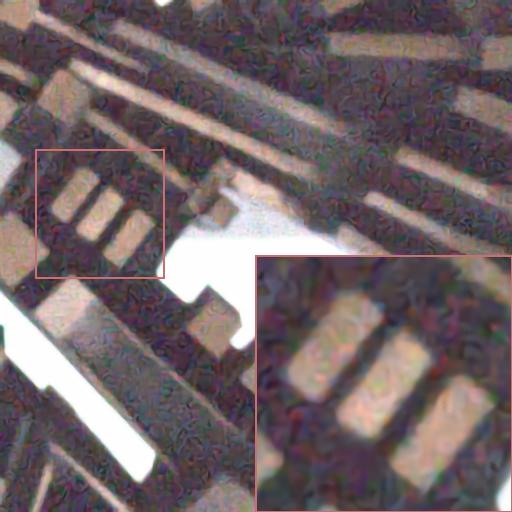}{}{28.66dB}{0.25,1.2}{ht}
      \tinyFigE{0.12}{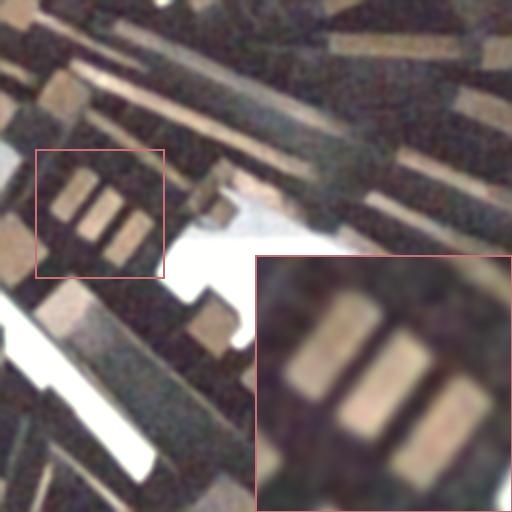}{}{30.37dB}{0.25,1.2}{ht}
      \tinyFigE{0.12}{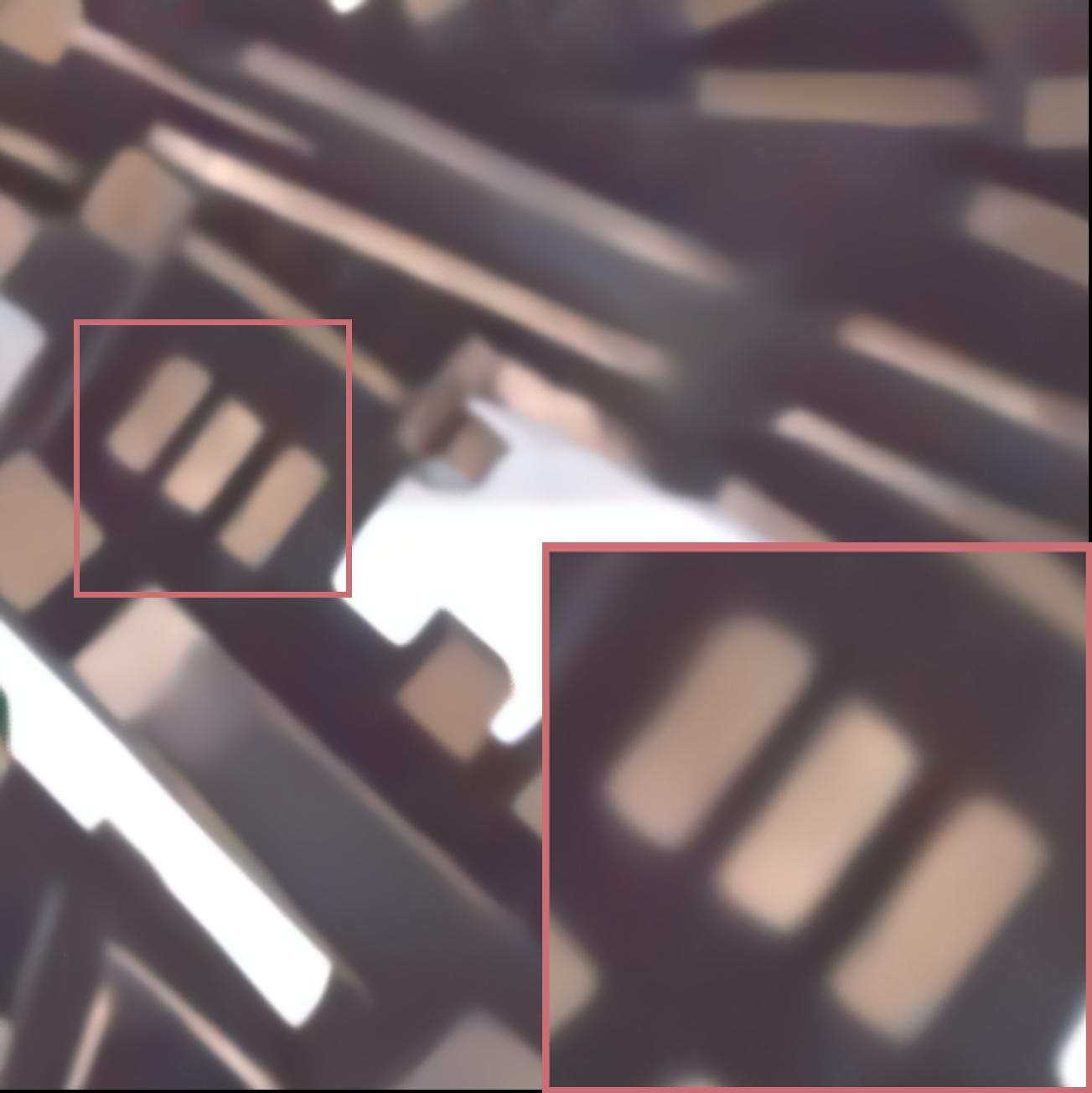}{}{33.81dB}{0.25,1.2}{ht}
      \tinyFigE{0.12}{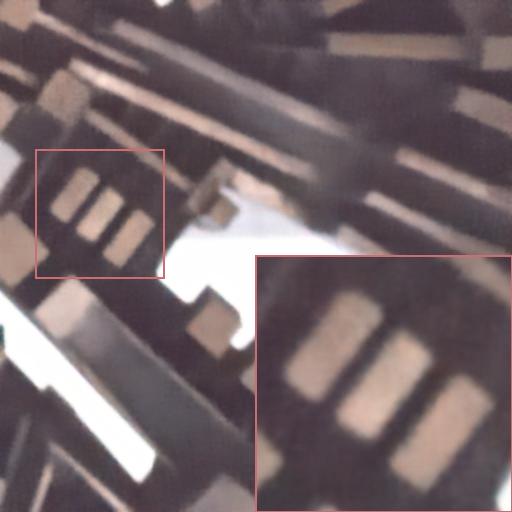}{}{33.49dB}{0.25,1.2}{ht}
      \tinyFigE{0.12}{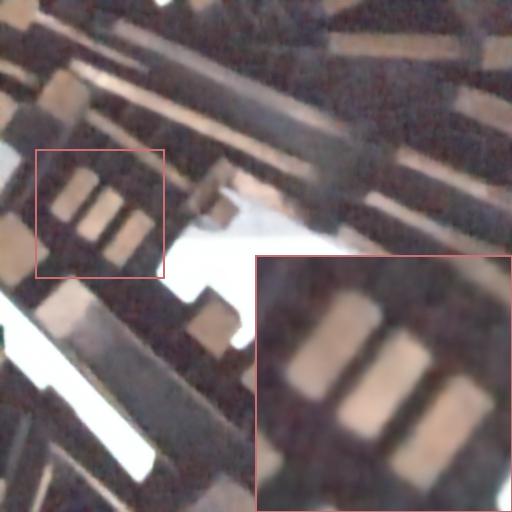}{}{32.63dB}{0.25,1.2}{ht}
      \tinyFigE{0.12}{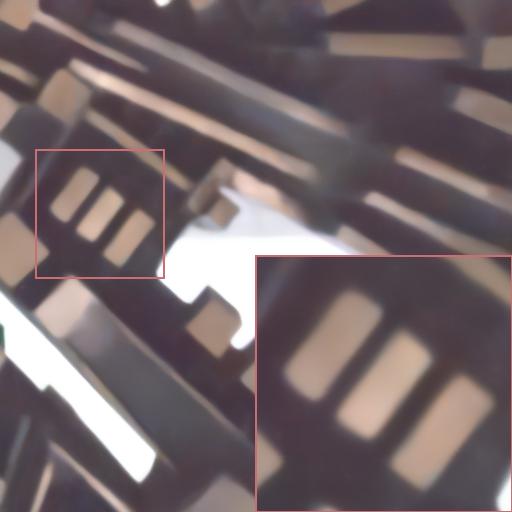}{}{34.19dB}{0.25,1.2}{ht}
      \caption{Visual comparison of our method against other denoising methods on the DND benchmark dataset. The image is (a) \textit{Noisy} (b) \textit{AP-BSN}\cite{lee2022apbsn} (c) \textit{CVF-SID}\cite{neshatavar2022cvf} (d) \textit{LG-BPN}\cite{wang2023lg} (e) \textit{BNN-LAN}\cite{li2023spatially} (f) \textit{SCPGabN}\cite{Lin_2023_ICCV} (g) \textit{\N-P-E}.}
      \label{fig:fig_0}
\end{figure}

\begin{table}[]
    \centering
    \setlength\tabcolsep{1pt}
    \renewcommand{\arraystretch}{1.4}
    \fontsize{6}{6}\selectfont
    \caption{Quantitative on PolyU with Restormer \cite{zamir2022restormer} as denoiser in \N. $\mathcal{R}^3$ denotes random refinement enhancement strategy.}
    \begin{tabular}{ccccccc}
        \toprule
        Method    & AP-BSN+$\mathcal{R}^3$ \cite{lee2022apbsn} & LG-BPN+$\mathcal{R}^3$ \cite{wang2023lg} & SCPGabN \cite{Lin_2023_ICCV} & BNN-LAN \cite{li2023spatially} & \textbf{\N-P} & \textbf{\N-P-E}       \\
        \midrule
        PSNR/SSIM & 36.88/0.9496                               & 36.25/0.9473                             & 37.11/0.9561                 & 37.13/0.9541                   & 37.47/0.9580  & \textbf{37.92/0.9645} \\
        \bottomrule
    \end{tabular}
    \label{table:polyu}
\end{table}


\subsection{Denoising of Real Images}

\subsubsection{Denoising on Public Real Noisy Datasets}
Our method focuses on self-supervised real-world denoising. \cref{table:Quantitative} compares the performance of various denoising method, on widely used SIDD and DND datasets. By using the enhancement strategies we introduced, our method \textbf{\N-P-E} achieves the state-of-the-art (SOTA). The visualization of some denoised images are shown in the \cref{fig:fig_47} and \cref{fig:fig_0}. More results can be found in the Supplementary Materials. For the restoration results of different self-supervised methods, CVF-SID \cite{neshatavar2022cvf} exhibits noticeable distortions, and SCPGabN \cite{Lin_2023_ICCV} has an overall inferior visual quality. The restoration results of AP-BSN \cite{lee2022apbsn}, BNN-LAN \cite{li2023spatially} and LGBPN \cite{wang2023lg} are all inferior to our method.

\cref{table:polyu} compares our method with several self-supervised methods on PolyU validation and our method \textbf{\N-P-E} achieves a best denoising performance. The more visualization results in Supplementary Materials.

\subsubsection{Validation on Self-Captured Real Noisy Images}
For real-world captured images denoising, we use the trained models \cite{neshatavar2022cvf,lee2022apbsn,li2023spatially,Lin_2023_ICCV} for fair comparsions. \cref{fig:canon_fig_0} compares the denoising results of different self-supervised methods on real noisy images captured by camera. In comparison to other methods, our approach provides a better visual quality, resulting in more natural denoised images with fewer distortions. CVF-SID \cite{neshatavar2022cvf} and AP-BSN \cite{lee2022apbsn} exhibit noticeable distortions and color blocks, while BNN-LAN \cite{li2023spatially} and SCPGabN \cite{Lin_2023_ICCV} have blurrier edges. More visualization results are in the Supplementary Materials.

\subsection{Ablation Work}

The ablation studies include identity mapping removal, denoiser selection, the effect of $\mathcal{L}_t$ and the effect of mask proportions on denoising results.

\label{sec:ablation1}
\textbf{Ablation study on identity mapping removal.} \cref{table:identity} demonstrates that without restricting the receptive field, our method circumvents the limitations inherent in conventional BSNs optimization strategies and avoids the risk of noise identity mapping in \cref{sec:reap}.

We benchmark against AP-BSN \cite{lee2022apbsn}, a representative BSN method for real-world denoising tasks and utilizing a denoiser with a limited receptive field after blind convolution, denoted as $D_A$. Substituting the dilated convolutions in $D_A$ with standard convolutions yields a variant named $D_{AS}$ which matches $D_A$ in parameter count and computational complexity. When using $D_A$ as the denoiser for \N\ and APBSN with corresponding self-supervised optimization strategy, both methods show great self-supervised denoising effects. However, with $D_{AS}$ as the denoiser, the AP-BSN framework succumbs to identity mapping, while our \N\ maintains its denoising effectiveness. Compared with the BSN-type strategies, our approaches does not require strict constraints on the denoiser design and can avoid the occurrence of identity mapping.
\begin{table}[htbp]
    \centering
    \fontsize{9}{9}\selectfont
    \setlength\tabcolsep{3pt}
    \renewcommand{\arraystretch}{1.4}
    \caption{AP-BSN \cite{lee2022apbsn} undergoes an identity mapping from noise to noise, yet \N\ retains its denoising capability. Each method employs random replacement refinement strategy.}

    \begin{tabular}{ccccc}
        \toprule
        Scheme/Denoiser & \N($D_A$)    & \N($D_{AS}$) & AP-BSN($D_{A}$) & AP-BSN($D_{AS}$) \\
        \midrule
        SIDD Validation & 37.25/0.8871 & 37.11/0.8863 & 36.74/0.8501    & 20.91/0.2447     \\
        SIDD Benchmark  & 37.17/0.934  & 37.08/0.933  & 36.91/0.931     & 21.93/0.253      \\
        \bottomrule
    \end{tabular}
    \label{table:identity}
\end{table}

\textbf{Ablation study on denoiser select.} \cref{sec:method} proposes that our scheme can freedom denoiser choice. To demonstrate it, we selecte five denoising methods as denoisers within our \N\ and train these denoisers on the SIDD Medium dataset. Subsequently, we validate their denoising performance on the SIDD validation and benchmark. Denoising results are shown in the \cref{table:ablation}. Our approaches allows the freedom to choose denoisers from different methods.

\begin{table}[htbp]
    \centering
    \fontsize{9}{9}\selectfont
    \renewcommand{\arraystretch}{1.4}
    \setlength\tabcolsep{1pt}
    \caption{Performance of different denoisers within \N.}
    \begin{tabular}{cccccc}
        \toprule
        Denoiser        & Restormer \cite{zamir2022restormer} & DeamNet \cite{ren2021adaptive} & DnCNN \cite{zhang2017beyond} & NAFNet \cite{chen2022simple} & UNet \cite{ronneberger2015u} \\
        \midrule
        SIDD Validation & 37.93/0.8948                        & 37.80/0.8923                   & 36.93/0.8864                 & 37.10/0.8814                 & 36.94/0.8854                 \\
        SIDD Benchmark  & 37.87/0.941                         & 37.71/0.937                    & 36.85/0.930                  & 37.08/0.934                  & 36.88/0.931                  \\
        \bottomrule
    \end{tabular}
    \label{table:ablation}
\end{table}

\begin{figure}[htbp]
    \centering
    \begin{minipage}[t]{0.36\textwidth}
        \centering

    \begin{tikzpicture}
        \begin{axis}[
                ybar,
                width=0.95\textwidth,
                height=1.1\textwidth,
                ymin=36.55,
                ymax=36.72,
                xtick={1,2},
                xmin=0.9,xmax=2.2,
                xticklabels={\shortstack{SIDD\\Validation}, \shortstack{SIDD\\Benchmark}},
                ylabel={PSNR (dB)},
                ylabel near ticks,
                enlarge x limits=true,
                enlarge y limits=true,
                legend style={at={(0.7,0.96)},
                        anchor=north,},
                yticklabel style={font=\small},
                xticklabel style={font=\small},
            ]
            \addplot coordinates {(1, 36.60) (2, 36.60)};
            \addplot coordinates {(1, 36.71) (2, 36.66)};
            \legend{$\mathcal{L}_m$ , $\mathcal{L}_t$ }
        \end{axis}
    \end{tikzpicture}
        \label{fig:loss_ablation}
        \caption{Denoising results with different losses with Restormer as denoiser.}
    \end{minipage}
    \begin{minipage}[t]{0.57\textwidth}
        \centering
        \begin{tikzpicture}
        \begin{axis}[
                width=0.95\textwidth,
                height=0.8\textwidth,
                xlabel={Mask Ratio},
                ylabel={PSNR (dB)},
                ymin=30, ymax=39,
                grid=both,
                grid style={line width=.1pt, draw=gray!10},
                major grid style={line width=.2pt,draw=gray!50},
                minor tick num=5,
                enlarge x limits=true,
                enlarge y limits=true,
                legend style={at={(0.8,0.3)}, anchor=north,legend columns=-1},
                xtick=data,
                xticklabels={75\%, 50\%, 25\%, 12.5\%, 6.25\%},
                yticklabel style={font=\small},
                xticklabel style={font=\small},
                ylabel near ticks,
                xlabel near ticks,
                every axis plot/.append style={thick},
            ]

            \addplot[mark=*,blue, style=dashed, mark options={solid}] coordinates {
                    (1, 34.12)
                    (2, 37.93)
                    (3, 37.90)
                    (4, 37.90)
                    (5, 35.83)
                };
            \addlegendentry{PSNR}

            \addplot[mark=*,blue, only marks, nodes near coords, point meta=explicit symbolic, nodes near coords style={above=1mm}] coordinates {
                    (1, 34.12) [34.12]
                    (2, 37.93) [37.93]
                    (3, 37.90) [37.90]
                    (4, 37.90) [37.90]
                    (5, 35.83) [35.83]
                };

        \end{axis}

        \begin{axis}[
                width=0.95\textwidth,
                height=0.8\textwidth,
                axis y line*=right,
                axis x line=none,
                ylabel={SSIM},
                ymin=0.85, ymax=0.93, 
                legend style={at={(0.4,0.3)}, anchor=north,legend columns=-1},
                ylabel near ticks,
                yticklabel style={font=\small, color=red},
                every axis plot/.append style={thick},
            ]

            \addplot[mark=square*,red, style=dotted, mark options={solid}] coordinates {
                    (1, 0.8589)
                    (2, 0.8948)
                    (3, 0.8940)
                    (4, 0.8941)
                    (5, 0.8817)
                };
            \addlegendentry{SSIM}

            \addplot[mark=square*,red, only marks, nodes near coords, point meta=explicit symbolic, nodes near coords style={above=1mm}] coordinates {
                    (1, 0.8589) [0.8589]
                    (2, 0.8948) [0.8948]
                    (3, 0.8940) [0.8940]
                    (4, 0.8941) [0.8941]
                    (5, 0.8817) [0.8817]
                };

        \end{axis}
    \end{tikzpicture}
        \captionsetup{width=.8\linewidth}
        \caption{Denoising results on the SIDD validation dataset with various mask ratios for each denoising branches.}
        \label{fig:differ_mask}
    \end{minipage}
\end{figure}
\textbf{Ablation study on loss function.} \cref{fig:loss_ablation} presents the effects of using $\mathcal{L}_m$ and $\mathcal{L}_t$ loss functions. With the introduction of $\mathcal{L}_t$ during the fine-tuning phase, the denoise performance has been improved about 0.1dB, which proves that introducing $\mathcal{L}_t$ loss can positively improve the reconstruction quality.

\textbf{Ablation study on different mask ratio.} We investigate the effect of different mask ratios on each denoising branch, as illustrated in \cref{fig:differ_mask}. All models are trained and validated under same mask ratios. When the mask proportion in each branch is approximately 50\%, we achieve the best denoising noise performance. This is also the rationale for set the number of denoising branches  $k$ to 2. See the Supplementary Materials for more details.

\begin{figure}
    \centering
    \begin{tikzpicture}
        \begin{axis}[
                width=0.8\textwidth,
                height=0.3\textwidth,
                xlabel={$\lambda$},
                ylabel={PSNR (dB)},
                ymin=36.4, ymax=36.8,
                major grid style={line width=.2pt,draw=gray!50},
                minor tick num=5,
                enlarge x limits=true,
                enlarge y limits=true,
                legend style={at={(0.8,0.3)}, anchor=north,legend columns=-1},
                xtick=data,
                xticklabels={1,0.5,0.2,0.1,0.05,0.01,0.001,0.0001},
                yticklabel style={font=\small},
                xticklabel style={font=\small},
                ylabel near ticks,
                xlabel near ticks,
                every axis plot/.append style={thick},
            ]

            \addplot[mark=*,blue, style=dashed, mark options={solid}] coordinates {
                    (1, 36.40)
                    (2, 36.45)
                    (3, 36.55)
                    (4, 36.71)
                    (5, 36.71)
                    (6, 36.62)
                    (7, 36.61)
                    (8, 36.61)
                };
            \addlegendentry{PSNR}

            \addplot[mark=*,blue, only marks, nodes near coords, point meta=explicit symbolic, nodes near coords style={above=1mm}] coordinates {
                    (1, 36.40) [36.40]
                    (2, 36.45) [36.45]
                    (3, 36.55) [36.55]
                    (4, 36.71) [36.71]
                    (5, 36.71) [36.71]
                    (6, 36.62) [36.62]
                    (7, 36.61) [36.61]
                    (8, 36.61) [36.61]
                };

        \end{axis}
    \end{tikzpicture}
    \caption{Denoising results on the SIDD validation dataset with different $\lambda$.}
    \label{fig:weight_ablation}
\end{figure}
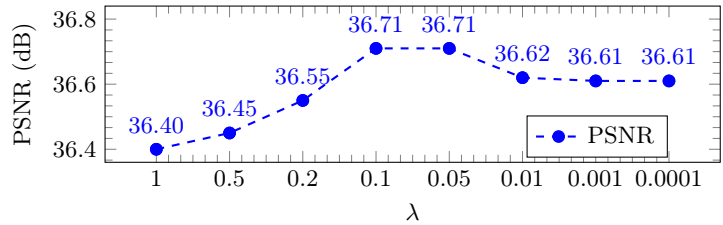

\textbf{Ablation study on $\lambda$.} We investigate the effect of $\lambda$, the weight coefficient of priori smoothing loss. As illustrated in \cref{fig:weight_ablation}, 0.1 is a good choice for high performance.
\section{Conclusion}
\label{sec:conclusion}
In this paper, we first analyze the reasons behind the limited performance of Blind Spot Network (BSN) when applied to self-supervised denoising. Inspired by Masked Autoencoders (MAE), we propose a mask-based self-supervised strategy to overcome the structural limitations inherent in BSN-type methods. We introduce the aymmetric mask schemes, which employs different operations during training and inference, to achieve expedited optimization and denoising for the entire noisy image. Through further analysis, we propose efficient strategies to enhance the final denoising performance. With our approach, the limitations of denoiser have been removed. According to introduce advanced denoisers, our \N\ achieves state-of-the-art denoising results. We believe our approach can offer valuable insights for various self-supervised real-world denoising techniques.

\noindent \textbf{Acknowledgements.} This work was supported by the National Natural Science Foundation of China under Grant 62171304, the Natural Science Foundation of Sichuan Province under Grant 2024NSFSC1423, and the Cooperation Science and Technology Project of Sichuan University and Dazhou City under Grant 2022CDDZ-09.

%
%
\bibliographystyle{splncs04}
\bibliography{main}
\end{document}